\documentclass[
reprint,
superscriptaddress,
showpacs,
amsmath,
amssymb,
aps,
prl,
floatfix,
]{revtex4-1}
\usepackage{graphicx} % Include figure files
\usepackage{dcolumn} % Align table columns on decimal point
\usepackage{bm} %bold math
\usepackage[colorlinks=true,urlcolor=blue,citecolor=blue,linkcolor=red]{hyperref}
\usepackage{color}
\usepackage{url}
\usepackage{braket}
\usepackage{subcaption}
\usepackage{hyperref}

\newcommand{\figref}[1]{Fig.~\ref{#1}}

\usepackage[english]{babel}
\usepackage{letltxmacro}

\LetLtxMacro{\ORIGselectlanguage}{\selectlanguage}
\makeatletter
\DeclareRobustCommand{\selectlanguage}[1]{%
  \@ifundefined{alias@\string#1}
    {\ORIGselectlanguage{#1}}
    {\begingroup\edef\x{\endgroup
       \noexpand\ORIGselectlanguage{\@nameuse{alias@#1}}}\x}%
}
\newcommand{\definelanguagealias}[2]{%
  \@namedef{alias@#1}{#2}%
}
\makeatother

\definelanguagealias{en}{english}
\begin{document}

\title{Fault-tolerant quantum computation with non-deterministic entangling gates}

\author{James M.~Auger}
\email{james.auger.09@ucl.ac.uk}
\affiliation{
 Department of Physics and Astronomy,
 University College London,
 Gower Street,
 London,
 WC1E 6BT,
 UK
}
\author{Hussain Anwar}
\affiliation{
 Department of Physics,
 Imperial College London,
 London,
 SW7 2AZ,
 UK
}
\affiliation{
 Department of Physics and Astronomy,
 University College London,
 Gower Street,
 London,
 WC1E 6BT,
 UK
}

\author{Mercedes Gimeno-Segovia}
\affiliation{Quantum Engineering Technology Labs, H. H. Wills Physics Laboratory and Department of Electrical and Electronic Engineering, University of Bristol, BS8 1FD, UK}
\affiliation{Institute for Quantum Science and Technology, University of Calgary, Alberta T2N 1N4, Canada}
\affiliation{
 Department of Physics,
 Imperial College London,
 London,
 SW7 2AZ,
 UK
}

\author{Thomas M.~Stace}
\affiliation{
 ARC Centre for Engineered Quantum Systems,
 University of Queensland,
 Brisbane 4072,
 Australia
}
\author{Dan E.~Browne}
\affiliation{
 Department of Physics and Astronomy,
 University College London,
 Gower Street,
 London,
 WC1E 6BT,
 UK
}

\begin{abstract}

Performing entangling gates between physical qubits is necessary for building a large-scale universal quantum computer, but in some physical implementations---for example, those that are based on linear optics or networks of ion traps---entangling gates can only be implemented probabilistically. In this work, we study the fault-tolerant performance of a topological cluster state scheme with local non-deterministic entanglement generation, where failed entangling gates (which correspond to bonds on the lattice representation of the cluster state) lead to a defective three-dimensional lattice with missing bonds. We present two approaches for dealing with missing bonds; the first is a non-adaptive scheme that requires no additional quantum processing, and the second is an adaptive scheme in which qubits can be measured in an alternative basis to effectively remove them from the lattice, hence eliminating their damaging effect and leading to better threshold performance. We find that a fault-tolerance threshold can still be observed with a bond-loss rate of 6.5\% for the non-adaptive scheme, and a bond-loss rate as high as 14.5\% for the adaptive scheme.

\end{abstract}
\maketitle

There are many current experimental proposals for building a universal quantum computer, and all of these suffer from the accumulation of errors that arise from the decoherence of physical quantum operations; these errors can be handled using standard quantum error correction codes. Some implementations---such as those that utilize optical components in constructing large-scale linear optical architectures~\cite{Knill2001-kx,*Browne2005-ko,*Rudolph2017-cc,Gimeno-Segovia2015-xz} or networks of trapped ions~\cite{HRB08,Nickerson2014-mr}---suffer from an additional problem in the form of non-deterministic entangling operations, a problem that has not been widely studied.

\begin{figure}
  \includegraphics[width=1\linewidth]{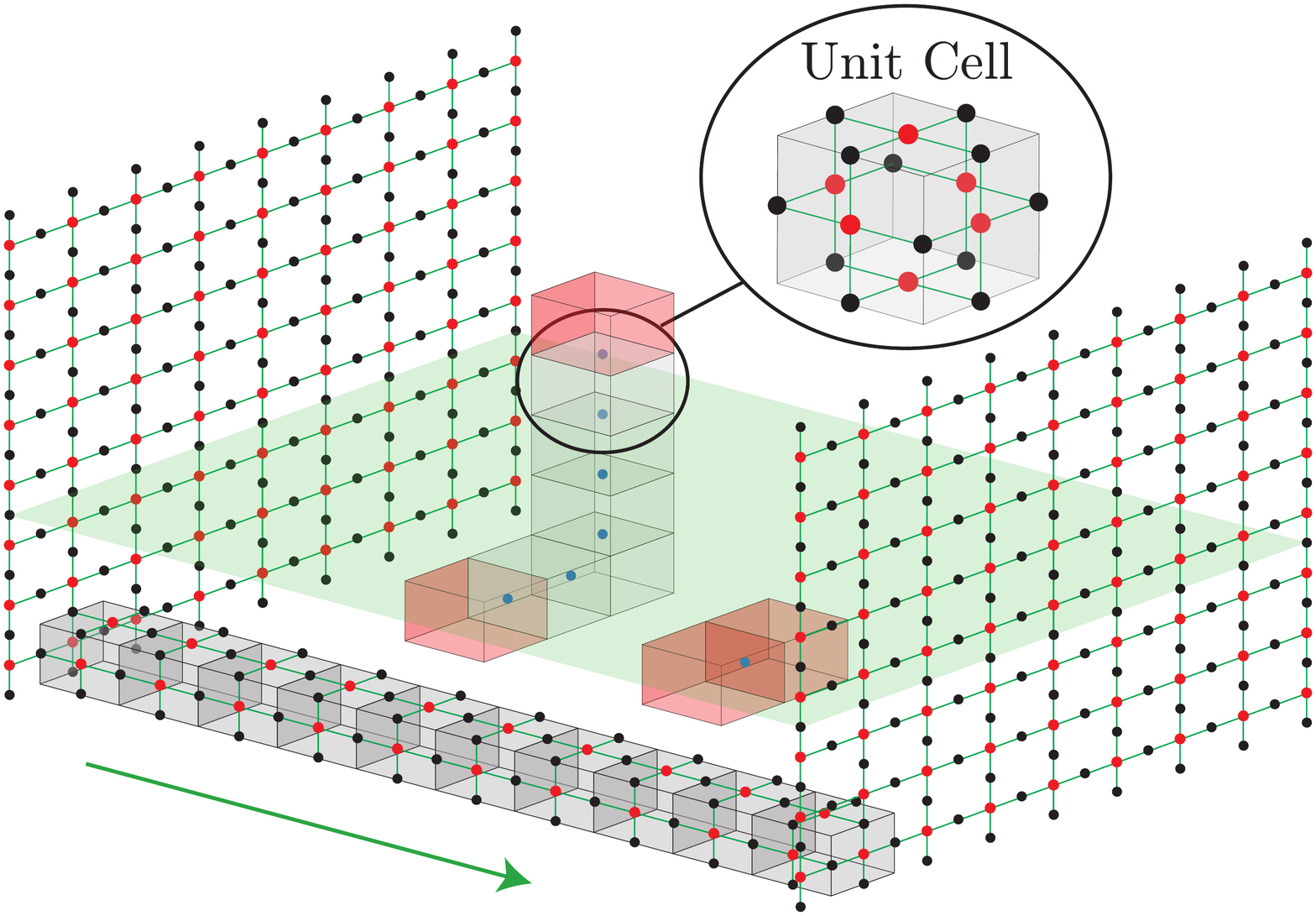}
  \caption{\label{fig:front_page}The three-dimensional topological cluster state, constructed from cubic cells (inset); bulk qubits are hidden for clarity. Only the ends of strings of $Z$ errors (blue qubits surrounded by wire-frame cubes) are detected by check operators (highlighted cubes). The correlation surface (shaded surface) spans the lattice in the direction of the computation (shown by the arrow).\vspace{-0.5cm}}
\end{figure}

In this Letter, we show that it is possible to perform fault-tolerant quantum computation with probabilistic entangling gates using the well-established topological cluster state scheme due to Raussendorf \emph{et~al.}~\cite{Raussendorf2006-ub,*Raussendorf2007-ih}---a three-dimensional measurement-based scheme that supports topological error correction. The scheme involves preparing qubits in a lattice configuration constructed with the unit cell shown in \figref{fig:front_page}. Entanglement is created between carefully chosen neighboring qubits during initialization, and error correction and quantum computation then proceed by single-qubit operations alone with no further multi-qubit operations~\cite{Raussendorf2006-ub,*Raussendorf2007-ih}. Topological protection is achieved by having the surface code as a substrate at each layer of the cluster state, such that the two-dimensional logical operators of each surface code are extruded into to the third dimension to form \emph{correlation surfaces} that encode logical information globally.

\begin{figure}
  \includegraphics[width=1\linewidth]{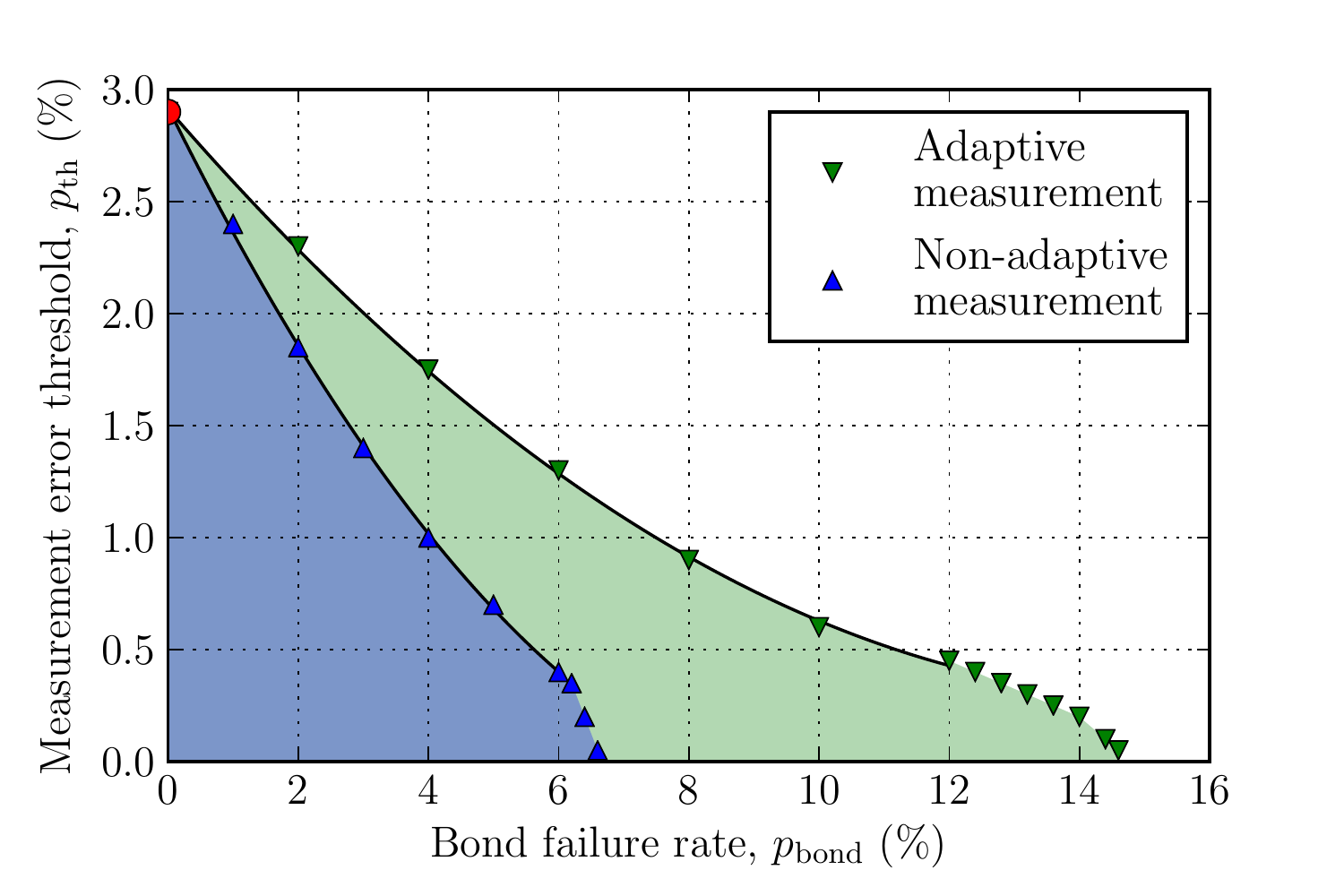}
  \caption{\label{fig:cluster_thresholds}Thresholds in the presence of failed bonds. The shaded regions indicate correctable error rate combinations. In the absence of bond failures ($p_\mathrm{bond}=0$), the threshold (shown by the marker on the y-axis) agrees with~\cite{Wang2003-mh}.\vspace{-0.5cm}}
\end{figure}

We propose two approaches for handling non-deterministic entanglement generation in Raussendorf's scheme: a non-adaptive approach, which involves the same measurement pattern as the original scheme~\cite{Raussendorf2006-ub,*Raussendorf2007-ih} with no additional quantum processing, and an adaptive approach, which involves changing the basis in which some qubits are measured. Our main result is shown in \figref{fig:cluster_thresholds}, which shows the quantum error correction thresholds obtained from simulations of both approaches.

Our primary motivation for this work comes from linear optical architectures~\cite{Knill2001-kx,*Browne2005-ko,*Rudolph2017-cc,Gimeno-Segovia2015-xz}, but our analysis is sufficiently general that the qualitative results are relevant to other implementations with non-deterministic entanglement. The approach we describe relaxes the need for deterministic or repeat-until-success~\cite{Lim2006-yd} entanglement generation, and we show that Raussendorf's scheme can tolerate a degree of failure in the construction of the underlying cluster state bonds.

Previous work along similar lines of research include~\cite{Barrett2010-sz,Whiteside2014-nw, SBD09}, which considered qubit loss and leakage in topological codes,~\cite{YBS10}, which considered the construction of topological codes with non-deterministic entanglement between multi-qubit resource states, and~\cite{Fowler2010-la}, which considered surface code based quantum repeaters with non-deterministic entanglement between nodes but deterministic entanglement within nodes. Our work differs from that of~\cite{YBS10} in that we are considering non-deterministic entanglement between all qubits, rather than between networks of multi-qubit nodes.

%%%%%TCS%%%%%%%%
\emph{Topological~cluster~states.}---In this work we discuss the cubic topological cluster state (TCS)~\cite{Raussendorf2006-ub,*Raussendorf2007-ih}, which consists of qubits in a lattice configuration based around the unit cell shown in \figref{fig:front_page}; the edges between qubits in this figure are referred to as \emph{bonds}. Entanglement is created between carefully chosen neighboring qubits during initialization to form a \emph{cluster state}~\cite{Briegel2001-pv}. The quantum state of the lattice is equivalent to that obtained by preparing every qubit in the $\ket{+}$ state and performing Controlled-$Z$ (CPHASE) gates between qubits linked by bonds. The lattice can be created in different but equivalent ways without explicit $\ket{+}$ state preparation and CPHASE gates, such as the linear optics scheme in~\cite{Gimeno-Segovia2015-xz}.

This lattice structure gives rise to primal and dual lattices that are used for error correction---two interleaved cubic lattices, one with the black qubits from \figref{fig:front_page} on the center of each cube face and the other with the red qubits on the center of each face. For each qubit, $i$, in a cluster state, there is an associated stabilizer operator, $S_i$, of the form
\begin{align}
    S_i = X_i \bigotimes_{j \in N(i)} Z_j,
\end{align}
where $N(i)$ is the neighborhood of qubit $i$ (the adjacent qubits). Therefore, for each cube face, $f_i$, centered on qubit $i$, there is an associated stabilizer generator $S_i$ with an $X$ operator acting on qubit $i$ and $Z$ operators acting on the adjacent qubits.

By multiplying the six face operators of each cube together, the $Z$ contributions cancel, leaving a six-body operator with $X$ operators acting on the qubit at the center of each face. Stabilizer measurements can therefore be performed by measuring each face qubit in $X$ and multiplying the outcomes to form a parity check associated with that cube; the term \emph{check operator} will be used to describe these parity checks.

In the absence of errors, all check operators have parity `$+1$'. If a $Z$ error or a measurement error (incorrect outcome) occur on a single qubit on one face of a cube, this flips the parity of the check operator associated with that cube from `$+1$' to `$-1$' and also flips the parity of the corresponding adjacent cube. Errors on qubits on two faces of a cube will not change the parity of that cube's check operator but will be detected by the two adjacent cubes, such that the check operators detect only the ends of error strings (see Fig.~\ref{fig:front_page}). Homologically trivial error strings---those that form closed loops---are equivalent to logical identity operations. Strings that span the lattice in an undetectable fashion result in uncorrectable logical errors. The length of the shortest undetectable string across the lattice gives the code distance, so a lattice with a shortest dimension of $n$ unit cells has code distance $n+1$.

Once all qubits are measured in the $X$ basis, an error syndrome is obtained by collating the check operator outcomes. This syndrome provides the locations of the ends of error strings on the primal and dual lattices, and the decoder attempts to pair these to find a correction that minimizes the probability of a logical error occurring.

One can consider a lattice where a surface code state is input at one end, the measurements are performed, and a corresponding surface code state is output at the other end~\cite{Raussendorf2005-do}. In this picture, one of the three dimensions plays a similar role to that of time for a surface code with repeated syndrome measurements. The primal and dual lattices of the TCS  then each have a \emph{correlation surface} linking the input and output surfaces, as shown in \figref{fig:front_page}. The combined parity of the measurement outcomes of qubits in each of the correlation surfaces indicates whether a logical Pauli correction is required on the output state to mitigate the effect of errors and compensate for random measurement outcomes. Each correlation surface can be deformed to a logically equivalent operator by multiplication with an element of the stabilizer group (i.e. a cube) such that the correlation surface is not unique.

\begin{figure}
  \includegraphics[width=1\linewidth]{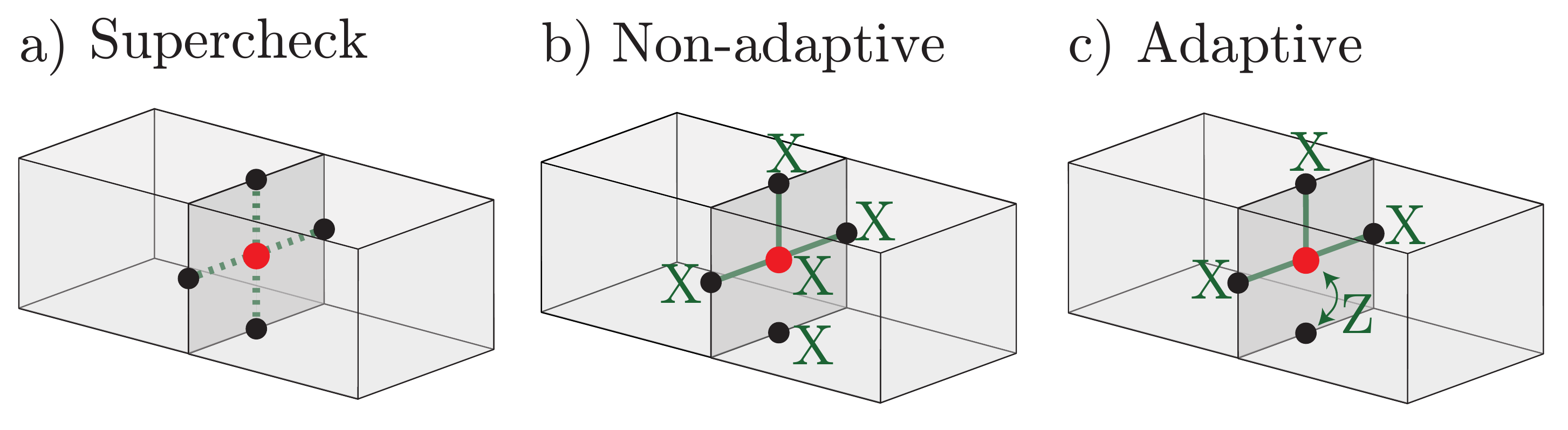}
  \caption{\label{fig:deformation}(a) Multiplying two cubes to form a supercheck removes the face qubit shared between them and results in a parity check involving the $X$ measurements associated with the ten remaining face qubits. (b) In the non-adaptive approach, all bulk qubits are measured in the $X$ basis in the presence of a failed bond. (c) In the adaptive approach, one of the qubits incident on the missing bond is randomly chosen to be measured in the $Z$ basis while the other is measured in the $X$ basis.\vspace{-0.5cm}}
\end{figure}

%%%%% BOND LOSS %%%%%
\emph{Bond loss.}---This work considers the impact of bond failures in TCS schemes, i.e.~when certain bonds between qubits are never created. Such errors are relevant to any TCS scheme where entangling operations can fail, particularly linear optics schemes using fusion gates~\cite{Gimeno-Segovia2015-xz}. Our results also provide insights for surface code schemes with non-deterministic two-qubit gates due to the similarity between TCS and surface codes.

A failed bond has a similar effect to losing the qubits at either end of a successful bond. Qubit loss in TCS was considered in~\cite{Barrett2010-sz}, which looked at TCS schemes in which all bonds were successful but some qubits were lost before and after bond creation. Lost qubits are handled during the error correction procedure by combining multiple cubes to form \emph{supercheck} operators made up of more than six measurement outcomes---whenever a qubit is lost, the two check operators associated with the adjacent cubes are multiplied together to remove the effect of the lost qubit from the parity check. An example of a supercheck is shown in \figref{fig:deformation}(a). This procedure restores the error correcting properties of the code at the cost of reduced code distance, and tolerates qubit loss rates up to 24.9\%.

Later work by \cite{Whiteside2014-nw} expanded this analysis by considering a gate-based TCS scheme experiencing dynamic loss during all stages of the computation, not just initialization and measurement. This analysis resulted in a higher effective loss rate per qubit, and correspondingly lower loss threshold of 2-5\% \emph{per operation} (rather than per qubit).

Our work mitigates the effect of failed bonds using a similar procedure to~\cite{Barrett2010-sz}. To isolate the impact of failed bonds, it is assumed that qubit loss does not occur, and it is assumed that the locations of all failed bonds are known---we refer to this as \emph{heralded} bond loss. We propose two approaches for dealing with failed bonds. In the first method, called the non-adaptive method, every bulk qubit is measured in $X$ as normal (see Fig.~\ref{fig:deformation}(b)). In the second method, called the adaptive method, certain qubits are measured in the $Z$ basis to remove them from the lattice (see Fig.~\ref{fig:deformation}(c)). It should be noted that both approaches can also handle qubit loss, in which case there will be a trade-off between tolerable qubit loss rates and bond failure rates.

In the non-adaptive method, bond failures are mapped onto the qubits by treating the qubit at each end of the bond as a lost qubit in the picture of~\cite{Barrett2010-sz}; this means that all additional processing is performed classically during decoding and no extra quantum resources are required. Each bond touches two qubits: one on the primal and one on the dual lattice. Without loss of generality, we consider qubits on the primal lattice. When a bond fails, the associated qubit is removed from the error correction procedure by combining the two incident check operators. This process is repeated until all qubits involving failed bonds are removed. If such a qubit is part of the correlation surface, the correlation surface is modified by multiplication by an appropriate check or supercheck operator to remove the qubit from the correlation surface. If the removed qubits form a continuous string percolating the primal lattice such that a correlation surface cannot be formed, a percolation error has occurred and the code is uncorrectable. An analogous process can be performed for the dual lattice.

In the adaptive method, bond failures are mapped onto the qubits by measuring a qubit at one end of the bond in the $Z$ basis. The qubit to measure in $Z$ is chosen at random, and a qubit is only measured in $Z$ if the adjacent qubit has not already been measured in $Z$. In this case, the qubits that are measured in $Z$ are treated identically to lost qubits in~\cite{Barrett2010-sz}, and formation of superchecks and correlation surfaces proceeds in the same manner as the non-adaptive scheme except that each failed bond affects only one of the primal or dual lattice at random, not both. This adaptive approach leads to an improved threshold at the cost of requiring more quantum processing (i.e. the ability to change measurement basis during the computation). It should be noted that the adaptive approach remains a measurement-based quantum computing scheme without any additional entangling gates, and measuring qubits in the $Z$ basis is already required to perform quantum computation (although the locations of such $Z$ measurements is generally pre-determined).

%%%%% SIMULATIONS %%%%%
\emph{Simulations.}---Both methods are simulated to obtain error correction thresholds. In the simulations, each bond has a probability $p_{\mathrm{bond}}$ of failing; a failed bond is considered to have never existed. All bond failures occur independently and are heralded. Additionally, each measurement outcome has an independent probability $p_{\mathrm{comp}}$ of being incorrect. This model is chosen to give an indication of the effect of failed bonds without considering a specific implementation. For example, one could assign a Pauli error probability to each CPHASE gate when constructing the lattice, but such a model is not appropriate for the scheme in~\cite{Gimeno-Segovia2015-xz}, where CPHASE gates are not used. The chosen error model is qualitatively similar to the \emph{random-plaquette gauge model} used in~\cite{Wang2003-mh} for the toric code, and gives a similar threshold in the absence of failed bonds.

The simulations use lattices with a range of code distances $d$, with the first `input' layer acting as a surface code state. This layer is followed by $4d-2$ layers of qubits, finishing with an `output' surface code layer ($4d-1$ layers in total), giving the lattice a depth of $2d$ cubes. For simplicity, we assume that bonds involving only qubits in the first two or final two layers always succeed, and measurements of the black qubits in the first and final layers and red qubits in the second and penultimate layers are always perfect; this is to ensure that the code is projected into a valid surface code state at each end following all measurements, as is standard with many quantum error correction simulations (in reality far more than $4d$ layers would be involved in a computation).

Except for the $Z$ basis measurements in the adaptive scheme, all qubits in the bulk of the lattice are measured in the $X$ basis. This leaves the first and final layers in surface code states, with the encoded state forming a Bell pair if a percolation error does not occur~\cite{Raussendorf2005-do}. The order in which gates are performed and the order in which measurements are performed are unimportant for the chosen error model provided that gates precede measurements on any particular qubit; the simulation creates bonds in an arbitrary order and then performs all measurements in an arbitrary order, but it would be equally valid to alternate rounds of bond creation and measurement.

The decoder uses the knowledge of the locations of failed bonds and mappings outlined above to form superchecks, and a minimum-weight perfect matching algorithm~\cite{Raussendorf2006-ub,*Raussendorf2007-ih} is used to calculate the required correction, with the edge weights set using the methods in~\cite{Barrett2010-sz}. If the resulting logical Bell pair between the surface codes of the first and final layers is $\ket{\Phi^+}_{\mathrm{L}} = \frac{1}{\sqrt{2}}(\ket{00}_\mathrm{L} + \ket{11}_\mathrm{L})$, the error correction is deemed successful. If the resulting Bell pair is not $\ket{\Phi^+}_{\mathrm{L}}$ or a percolation error occurs, then the error correction is deemed unsuccessful and a logical error has occurred. For each value of $p_{\mathrm{bond}}$, a computational threshold, $p_{\mathrm{th}}$, is determined from the intersection point of logical error rates for code distances of 7, 9, 11 and 13~\cite{Wang2003-mh}.

%%%%% RESULTS/DISCUSSION %%%%%

\emph{Results.}---\figref{fig:cluster_thresholds} shows the threshold results of the simulations for the non-adaptive and adaptive methods. In the absence of failed bonds, the error model results in a threshold of $p_{\mathrm{th}} \approx 2.9\%$, in agreement with~\cite{Wang2003-mh}. The threshold decreases with increasing bond failure rates for both schemes; the non-adaptive scheme has a fit of $p_\mathrm{th} = 0.029 - 0.587 p_{\mathrm{bond}} + 2.786 p_{\mathrm{bond}}^2$ applied between $p_{\mathrm{bond}}=0\%$ and $p_{\mathrm{bond}}=6\%$, and the adaptive scheme has a fit of $p_{\mathrm{th}} = 0.029 - 0.336 p_{\mathrm{bond}} + 1.071 p_{\mathrm{bond}}^2$ between $p_{\mathrm{bond}}=0\%$ and $p_{\mathrm{bond}}=12\%$. The threshold for the non-adaptive and adaptive schemes disappears at $p_{\mathrm{bond}} \approx 6.5\%$ and $p_{\mathrm{bond}} \approx 14.5\%$, respectively; these limits are due to the percolation threshold for each method.

To link these results to those in~\cite{Barrett2010-sz}, we consider an approximate mapping from the $24.9\%$ percolation limit found for qubit loss (our error model results in an effective loss rate per qubit rather than per operation, so the threshold in~\cite{Barrett2010-sz} is more relevant to our analysis than that in~\cite{Whiteside2014-nw}). In the non-adaptive approach, each bulk qubit has four bonds, and qubits with failed bonds are treated equivalently to lost qubits in~\cite{Barrett2010-sz}. The probability of a bulk qubit having one or more failed bonds is $1 - (1-p_\mathrm{bond})^4$, resulting in an expected percolation threshold when $1 - (1-p_\mathrm{bond})^4 = 0.249$, or $p_\mathrm{bond} = 6.9\%$, which is close to the value obtained.

For the adaptive scheme, each failed bond is mapped to just one of the two adjacent qubits. Therefore, the probability of a qubit being measured in the $Z$ basis receives a contribution of $\frac{1}{2}p_\mathrm{bond}$ from each incident bond. The probability that a particular qubit with four attempted bonds is measured in $Z$ is therefore $1-(1-\frac{1}{2}p_{\mathrm{bond}})^4$, resulting in an expected percolation threshold when $1-(1-\frac{1}{2}p_{\mathrm{bond}})^4 = 0.249$, or $p_{\mathrm{bond}} = 13.8\%$. This is slightly lower than the value obtained in the simulations as it does not account for neighboring qubits that have already been measured in the $Z$ basis.

% The phenomenological bond loss model assumed in this Letter can be connected to the microcluster scheme in~\cite{Gimeno-Segovia2015-xz}, which generates large-scale cluster states from elementary three-qubit GHZ resource states by performing a sequence of probabilistic fusion gates. We have performed additional numerical simulations on a modified version of this microcluster scheme, where we construct a Raussendorf lattice instead of the brickwork lattice used in the original proposal. We have found the relationship between fusion-gate success rates and bond failure rates in the TCS scheme by calculating average bond failure rates when varying the fusion-gate success rate. The results of this simulation suggest that the adaptive scheme would require a fusion success rate in excess of 95\% to perform fault-tolerant quantum computation, and the non-adaptive scheme would require a fusion-gate success rate in excess of 98\%. The fusion-gate success rate can be increased to this level by using larger resources to perform the gate, and although this increases the resources \emph{per fusion gate}, it is not clear that this will increase the resources overall, as the scheme we present in this Letter would not require additional procedures to perform fault-tolerant quantum computation, unlike that in~\cite{Gimeno-Segovia2015-xz}.

The phenomenological bond loss model assumed in this Letter can be connected to the microcluster scheme in~\cite{Gimeno-Segovia2015-xz}, which generates large-scale cluster states from elementary three-qubit GHZ resource states by performing a sequence of probabilistic fusion gates. To do so, we have performed additional numerical simulations on a modified version of this microcluster scheme to find the relationship between fusion-gate success rates and bond failure rates. The simulation involves using fusion gates to repeatedly create TCS lattices (unlike the brickwork lattice used in the original linear optical proposal) with code distance $d=6$ and fusion-gate success rates ranging from 50\% to 99.5\% in steps of 0.5\%. The proportion of missing bonds is measured for each lattice, and this is then averaged over all runs for each fusion-gate success rate to give an effective bond failure rate for each fusion-gate success rate. The minimum  fusion-gate success rate required to perform fault-tolerant quantum computation for the adaptive and non-adaptive schemes is found by mapping back from effective bond failure rates of $14.5\%$ and $6.5\%$ respectively, which correspond to the bond failure limits of each scheme, to fusion-gate success rates.

The results of the simulation suggest that the adaptive scheme would require a fusion-gate success rate in excess of 95\% to perform fault-tolerant quantum computation, and the non-adaptive scheme would require a fusion-gate success rate in excess of 98\%. The fusion-gate success rate can be increased to this level by using larger resources to perform the gate, and although this increases the resources \textit{per fusion gate}, it is not clear that this will increase the overall resources required, as the scheme we present would not require additional procedures to perform fault-tolerant quantum computation, unlike that in~\cite{Gimeno-Segovia2015-xz}.
%%%%% SUMMARY/CONCLUSION %%%%%

\emph{Conclusion.}---We have shown that fault-tolerant quantum computation can be performed with TCS schemes for probabilistic heralded entangling gate failure rates as high as 14.5\% if adaptive measurements are allowed, or as high as 6.5\% with no additional quantum overhead. Our findings are particularly relevant to linear optics schemes, but our approach is sufficiently general that it can be applied to any system with non-deterministic entangling gates. The shared features of topological codes mean our results also give a qualitative insight for other topological codes, such as the surface code, with non-deterministic two-qubit gates, building on the recent work of~\cite{Auger2017-di}. Future work includes considering unheralded entanglement failure, where the locations of missing bonds are unknown, and combining this approach with~\cite{Nickerson2014-mr} to attempt to reduce qubit and time overheads due to probabilistic GHZ state distillation for networks of trapped ions.
% \end{document}

\begin{acknowledgments}
\emph{Acknowledgements.}---We thank Terry Rudolph, Pete Shadbolt and Sam Morley-Short for useful discussions, and we thank Naomi Nickerson for her contribution to the adaptive measurement approach. JMA and HA are supported by EPSRC. MGS is supported by EPSRC and ERC. TMS acknowledges funding from The Australian Research Council (ARC), and an ARC Future Fellowship. The authors acknowledge the use of the UCL Legion High Performance Computing Facility (Legion@UCL). Our simulations used \emph{Blossom~V}~\cite{Kolmogorov2009-md} and \emph{GraphSim}~\cite{Anders2006-gv}.
\end{acknowledgments}

\end{document}